\RequirePackage{etoolbox} 
\newtoggle{prb}

\toggletrue{prb} 

\iftoggle{prb}{
\documentclass[aps,prb,reprint, superscriptaddress,amsmath,showpacs, showkeys, floatfix,a4paper,balancelastpage
]{revtex4-1} 
}{
\documentclass[fp,twocolumn]{jpsj3}
}

\usepackage[latin1]{inputenc} 
\usepackage{graphicx}
\usepackage{color,soul}  
\usepackage{bm}

\newcommand {\dg} {\ensuremath{^{\circ}}}
\newcommand {\mub} {\ensuremath{\mu_{\text{B}}}}

\newcommand {\V}[1] {$\bm{#1}$} 
\newcommand {\etal} {\textit{et al.}}

\newcommand {\Tz} {\ensuremath{T_{0}}}
\newcommand {\TN} {\ensuremath{T_{N}}}
\newcommand {\CeRuAl} {CeRu$_{2}$Al$_{10}$}
\newcommand {\CeOsAl} {CeOs$_{2}$Al$_{10}$}

\newcommand {\NdFeAl} {NdFe$_{2}$Al$_{10}$}
\newcommand {\TbFeAl} {TbFe$_{2}$Al$_{10}$}
\newcommand {\YbFeAl} {YbFe$_{2}$Al$_{10}$}
\newcommand {\CeTAl} {Ce$T_{2}$Al$_{10}$}

\iftoggle{prb}{
\newcommand{\citen} {\onlinecite}
}{}

\begin{document}

\title{Detailed neutron diffraction study of magnetic order in \NdFeAl}

\iftoggle{prb}{  

\author{Julien Robert}
\author{Françoise Damay}
\affiliation{Laboratoire L\'{e}on Brillouin, CEA-CNRS, CEA/Saclay, 91191 Gif sur Yvette, France}

\author{Kotaro Saito}
\altaffiliation[Present address: ]{Institute of Materials Structure Science, KEK, 1-1 Oho, Tsukuba, Ibaraki 305-0801, Japan} 
\affiliation{Laboratoire L\'{e}on Brillouin, CEA-CNRS, CEA/Saclay, 91191 Gif sur Yvette, France}  

\author{Alexandre M. Bataille}
\author{Florence Porcher}
\author{Gilles André}
\author{Arsen Gukasov}
\author{Jean-Michel Mignot}
\email[e-mail address: ]{jean-michel.mignot@cea.fr}
\affiliation{Laboratoire L\'{e}on Brillouin, CEA-CNRS, CEA/Saclay, 91191 Gif sur Yvette, France}

\author{Hiroshi Tanida}
\author{Masafumi Sera}
\affiliation{Department of Quantum Matter, ADSM, Hiroshima University, Higashi-Hiroshima, 739-8530, Japan}

\date{\today}

\begin{abstract}
The orthorhombic compound \NdFeAl\ has been studied by powder and single-crystal neutron diffraction. Below $T_N = 3.9$ K, the Nd$^{3+}$ magnetic moments order in a double-\V{k} [$\bm{k}_1 = (0, \frac{3}{4}, 0)$, $\bm{k}_2 = (0, \frac{1}{4}, 0)$] collinear magnetic structure, whose unit cell consists of four orthorhombic units in the $b$ direction.The refinements show that this structure consists of (0\ 1\ 0) ferromagnetic planes stacked along $b$, in which the moments are oriented parallel to $a$ (the easy anisotropy axis according to bulk magnetization measurements) and nearly equal in magnitude ($\approx 1.7-1.9 \mub$). The alternating \hbox{8-plane} sequence providing the best agreement to the data turns out to be that which yields the lowest exchange energy if one assumes antiferromagnetic near-neighbor exchange interactions with $J_1 \gg J_2, J_3$. With increasing temperature, the single-crystal measurements indicate the suppression of the $\bm{k}_2$ component at $T = 2.7$ K, supporting the idea that the anomalies previously observed around 2--2.5 K result from a squaring transition. In a magnetic field applied along the $a$ axis, the magnetic Bragg satellites disappear at $H_c = 2.45$ T, in agreement with earlier measurements. Comparisons are made with related magnetic orders occurring in \CeTAl\ ($T$: Ru, Os) and \TbFeAl. 
\end{abstract}



\maketitle
}{   
\author{
Julien Robert$^1$,
Françoise Damay$^1$,
Kotaro Saito$^{1,2}$\thanks{Present address: Institute of Materials Structure Science, KEK, 1-1 Oho, Tsukuba, Ibaraki 305-0801, Japan},
Alexandre M. Bataille$^1$,
Florence Porcher$^1$,
Gilles André$^1$,
Arsen Gukasov$^1$,
Jean-Michel Mignot$^1$\thanks{jean-michel.mignot@cea.fr},
Hiroshi Tanida$^3$,
Masafumi Sera$^3$}

\inst{$^1$Laboratoire L\'eon Brillouin, CEA-CNRS, CEA/Saclay, 91191 Gif sur Yvette, France \\
$^2$Department of Physics, Tohoku University, Sendai, Miyagi 980-8571, Japan \\
$^3$Department of Quantum Matter, ADSM, Hiroshima University, Higashi-Hiroshima 739-8530, Japan}

\abst{
The orthorhombic compound \NdFeAl\ has been studied by powder and single-crystal neutron diffraction. Below $T_N = 3.9$ K, the Nd$^{3+}$ magnetic moments order in a double-\V{k} [$\bm{k}_1 = (0, \frac{3}{4}, 0)$, $\bm{k}_2 = (0, \frac{1}{4}, 0)$] collinear magnetic structure, whose unit cell consists of four orthorhombic units in the $b$ direction.The refinements show that this structure consists of (0\ 1\ 0) ferromagnetic planes stacked along $b$, in which the moments are oriented parallel to $a$ (the easy anisotropy axis according to bulk magnetization measurements) and nearly equal in magnitude ($\approx 1.7-1.9 \mub$). The alternating \hbox{8-plane} sequence providing the best agreement to the data turns out to be that which yields the lowest exchange energy if one assumes antiferromagnetic near-neighbor exchange interactions with $J_1 \gg J_2, J_3$. With increasing temperature, the single-crystal measurements indicate the suppression of the $\bm{k}_2$ component at $T = 2.7$ K, supporting the idea that the anomalies previously observed around 2--2.5 K result from a squaring transition. In a magnetic field applied along the $a$ axis, the magnetic Bragg satellites disappear at $H_c = 2.45$ T, in agreement with earlier measurements. Comparisons are made with related magnetic orders occurring in \CeTAl\ ($T$: Ru, Os) and \TbFeAl.}
} 


\maketitle

\section{Introduction}
Orthorhombic compounds with formula Ce$T_2$Al$_{10}$ ($T$: Ru, Os, Fe) have attracted considerable interest recently because they exhibit a very peculiar coexistence of long-range magnetic order and Kondo-insulator properties,\cite{Muro'09, Nishioka'09, Strydom'09} which has been ascribed to an unusually strong anisotropy of the hybridization between $4f$ and conduction electron states.\cite{Kondo'11,Tanida'12} The antiferromagnetic (AFM) order reported for \CeRuAl\ and \CeOsAl\ has a number of intriguing features such as the very large value of the ordering temperature, $T_0 = 27$ K, in comparison with other $RT_2$Al$_{10}$ compounds ($R$: Nd, Gd),\cite{Kobayashi'11, Kunimori'12} the direction of the AFM moment ($\bm{m}_{\textrm{AF}} \parallel c$) not complying with the single-ion anisotropy (easy $a$ axis), or the lack of transverse, ``$\chi_{\perp}$''-type, behavior in the temperature dependence of the magnetic susceptibility $\chi_a (T)$ for a field applied along $a \perp \bm{m}_{\textrm{AF}}$, etc. Neutron scattering experiments have clarified several aspects of the magnetic structure\cite{Robert'10, Khalyavin'10, Mignot'11} (AFM with propagation vector $\bm{k} = (0, 1, 0)$ and a strongly reduced Ce magnetic moment, on the order of 0.3 $\mu_B$) and dynamics\cite{Khalyavin'10, Robert'10, Robert'12} (spin gap and dispersive magnetic excitations forming below \Tz). The spectrum of magnetic excitations, in particular, points to a strong anisotropy of exchange interactions, which competes with the conventional crystal-field anisotropy.

In an attempt to set an appropriate reference for the exotic magnetism occurring in the Ce compounds, experiments have been carried out on their Nd counterparts, which are expected to exhibit ``normal-rare-earth'' properties, dominated by crystal-field (CF) effects and Ruderman-Kittel-Kasuya-Yosida (RKKY) exchange interactions. The results reported by Kunimori \etal \cite{Kunimori'12} confirmed this expectation,  and the anisotropic magnetization (easy $a$ axis) in the paramagnetic regime could be reproduced satisfactorily using a simple two-sublattice mean-field model, with a small CF splitting, imputable to the large distance from the Ce ion to its ligands, and a moderate anisotropy in the exchange interactions. Below the ordering temperature, $\TN = 3.9$ K, the ordered moments orient along the easy anisotropy axis, and their magnitude looks consistent with the values calculated for the Nd$^{3+}$ CF doublet ground state. The overall picture is thus that of weakly interacting integral-valence rare-earth ions. Interestingly, anomalies have been observed in specific heat,\cite{Kunimori'12} $\mu$SR,\cite{Adroja'13rev} and electrical resistivity\cite{Tanida'priv} measurement at temperatures close to 2--2.5 K, suggesting the possible existence of a second transition below \TN. However, no indication of the nature  
of that transition has been reported so far. Previous neutron diffraction experiments on the isostructural heavy-rare-earth compound \TbFeAl\ have revealed a sequence of magnetic phases consisting of a sine-wave modulation with a wave vector very close to $\bm{k} = (0, 4/5, 0)$ forming at \TN = 16.5 K, followed by a squaring of the structure below 11 K.\cite{Reehuis'00,Reehuis'03} 

\begin{table*} [t] 
\centering
\caption{\label{tab:crysparam}Nuclear structure parameters of \NdFeAl\ (\YbFeAl\ structure) at room temperature.}
Wyckoff positions are: $4c$ [Nd], $8d$ [Fe], $8g$ [Al(1), Al(2)], $8f$ [Al(3), Al(4)], $8e$ [Al(5)].\cite{Thiede'98} 
\vspace {6pt}\
\begin{tabular}{l c c c c c c}
\hline\hline\
                                  & Space group &    $a_0$ (\AA)    &     $b_0$  (\AA)   &    $c_0$ (\AA)    & $R$ factor & $\chi^2$\\
 this work                   & $Cmcm$       & 9.0148(7) & 10.2168(7) & 9.0793(7) &       3.31     &   2.89  \\
 X-rays\footnote{Ref.~\citen{Sera'13}} & $Cmcm$       & 9.0158(1) & 10.2182(1) &  9.0800(1) &                  &            \\
\end{tabular}

\begin{tabular}{l c c c c c c c c c c}
 \hline
                                 & $y_{\mathrm{Nd}}$    & $x_{\mathrm{Al(1)}}$ & $y_{\mathrm{Al(1)}}$ & $x_{\mathrm{Al(2)}}$ & $y_{\mathrm{Al(2)}}$ & $y_{\mathrm{Al(3)}}$ & $z_{\mathrm{Al(3)}}$ & $y_{\mathrm{Al(4)}}$ & $z_{\mathrm{Al(4)}}$ & $x_{\mathrm{Al(5)}}$\\
 this work                               &          0.1238(2)             &          0.2290(3)            &        0.3617(3)              &            0.3500(3)          &            0.1286(3)           & 0.1602(3)           &          0.5975(3)             &          0.3765(3)            &        0.0514(3)              &            0.2280(2)\\
X-rays\footnote{Ref.~\citen{Sera'priv}}            &          0.12428             &          0.22727            &        0.36085              &            0.34989          &            0.12843           & 0.16077          &          0.59844             &          0.37661            &        0.05077              &            0.22721\\
\hline\hline
\end{tabular}
\end{table*}

In the present work we have performed neutron diffraction experiments to characterize the magnetic structure of \NdFeAl, which could not be determined from magnetic measurements alone. At the base temperature of $T = 1.7$ K, the structure is found to be commensurate, with cell dimensions $a \times 4b \times c$, and can be derived from the AFM structure of \CeRuAl\ and \CeOsAl\ by a change in the moment direction and the introduction of periodic spin-discommensurations.

\section{\label{ss:experiments}Experiments}

\begin{figure}  
	\centering
	\includegraphics [width=1.00\columnwidth, angle=0] {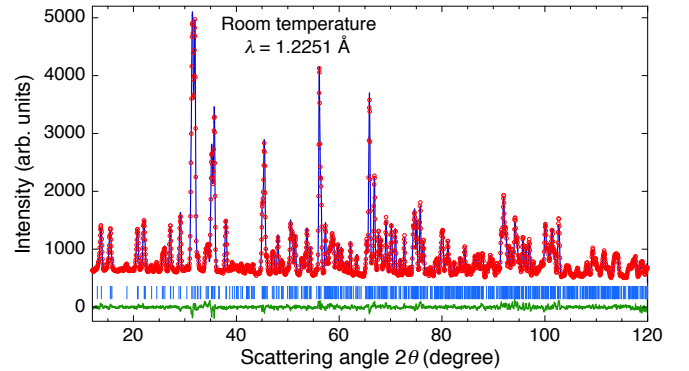}
	\caption{\label{diff_patt_3t2}(Color online) Rietveld refinement of the high-resolution neutron-diffraction pattern of \NdFeAl\ measured on 3T2 at room temperature. Open circles: measured intensities; full line through data: refinement; vertical marks: position of Bragg reflections; bottom trace: difference between measured and calculated intensities.}
\end{figure}

\NdFeAl\ was synthesized in single-crystal form starting from 99.9\% Nd and 99.995\% Fe constituents in a 99.999\% Al flux.
A powder sample of 6.3 g in mass was prepared by crushing small single-crystal pieces. Its quality was checked on the high-resolution powder diffractometer 3T2 (Orph\'ee-LLB), using an incident neutron wavelength $\lambda_i = 1.2251(2)$~\AA. The diffraction pattern is shown in Fig.~\ref{diff_patt_3t2}. The parameters derived from the Rietveld refinement, listed in Table~\ref{tab:crysparam}, are in excellent agreement with the space group  ($Cmcm$, \#63), lattice constants,\cite{Thiede'98,Sera'13} and atomic parameters \cite{Sera'13} determined previously by X-ray diffraction.
 
Neutron powder diffraction (NPD) patterns were collected on the two-axis diffractometer G4-1 (Orph\'ee-LLB, Saclay) equipped with a 800-cell position-sensitive detector. A monochromatic incident neutron beam of wavelength $\lambda_i =$2.426~\AA\ was produced by a pyrolytic graphite (PG002) monochromator, and higher-order contamination was suppressed by means of a PG filter. The sample powder was contained in a thin-walled cylinder-shape vanadium container, 6 mm in diameter, and cooled to 1.7 K in the exchange-gas chamber of a liquid-He cryostat. The data analysis was performed using the Rietveld refinement program \textsc{FullProf},\cite{fullprof'93,fullprof'01} with neutron scattering lengths and magnetic form factors taken from Refs.~\citen{Sears'92} and \citen{Freeman'79}, respectively. Absorption corrections were treated as negligible.

A large single-crystal ($m = 0.201$ g) was mounted on an aluminum sample holder with the $a$ axis vertical, and cooled to 1.7 K in the variable-temperature insert (VTI) of a 7-T, split-coil, vertical-field cryomagnet from Oxford Instruments. Single-crystal diffraction (SCD) experiments were performed on the thermal beam, two-axis neutron diffractometer Super-6T2 (Orph\'ee-LLB).\cite{Gukasov'07} 
Intensity maps were measured at two incident wavelengths, $\lambda_i = 0.902$~\AA\ (Cu monochromator, Er filter) and 2.345~\AA\ (PG002 monochromator, PG filter), by rotating the sample around the vertical axis with 
0.1\dg  
steps and recording the diffraction pattern in a \textsc{Bidim26} multiwire position-sensitive gas detector (PSD) developed at the ILL (Grenoble) and manufactured by INEL (Artenay, France). This procedure allowed us to explore a large three-dimensional (3D) segment of the reciprocal space by transforming a complete set of PSD images into the reciprocal space of the crystal. For quantitative refinements and studies of temperature and magnetic field dependences, the configuration was changed to a single lifting counter, with $20'$ Soller collimators. An extensive data set was collected at $\lambda_i = 0.902$~\AA\ in zero field at the base temperature, and more restricted ones at $T = 2.7$ K and at $H = 2.6$ T. The results were analyzed using the Cambridge Crystallography Subroutine Library (CCSL).\cite{ccsl'93}

\section{Results}

\subsection{\label{ss:magstrucpowd}Magnetic structure: powder diffraction}

\begin{figure}  
	\centering
	\includegraphics [width=0.97\columnwidth, angle=0] {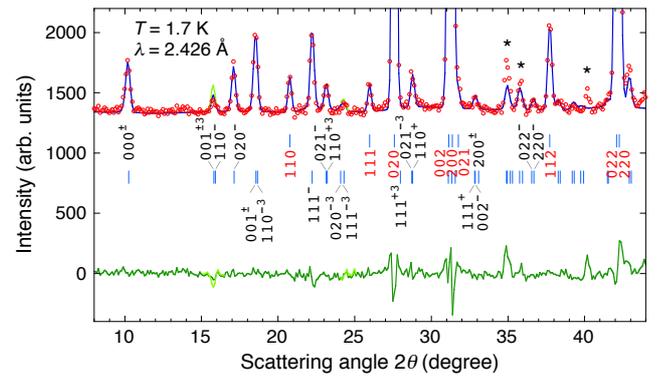}
	\caption{\label{diff_patt_1p7}(Color online)  Rietveld refinement of the neutron-diffraction pattern of \NdFeAl\ measured on G4-1 at $T = 1.7$ K. Open circles: measured intensities; full line through data: refinement using either the primitive ($P$), single-$\bm{k}$, or the centered ($C$), double-$\bm{k}$ representation of the structure (see text); vertical marks: position of Bragg reflections; bottom trace: difference between measured and calculated intensities; light-green peak near 15.5\dg: fit variant obtained by enforcing full squaring; stars signal positions at which contamination by a nonmagnetic impurity is observed for $T > \TN$.}
\end{figure}

The NPD pattern measured on G4-1 for $T = 1.7$ K is presented in Fig.~\ref{diff_patt_1p7}. Magnetic superstructure peaks are clearly visible at low scattering angles. However, these peaks cannot be indexed in the original $Cmcm$ space group using a single \V{k} vector. The reason (made obvious from the positions of the Bragg spots in the single-crystal map, to be discussed in the following) is that pairs of satellites occur not only near allowed nuclear reflections but also, with even larger intensities, near positions which are forbidden by the centering of the unit cell. If one assumes a propagation vector $\bm{k}_P = (0, \frac{1}{4}, 0)$, this would require treating the magnetic superstructure in the primitive orthorhombic lattice, although the nuclear reflections, measured in the same temperature range, give no indication of a symmetry lowering with respect to the $Cmcm$ space group. The latter was unambiguously assigned from accurate X-ray diffraction measurements at room temperature,\cite{Thiede'98,Sera'13} and is confirmed for the present sample by the neutron data collected on 3T2 (Sect.~\ref{ss:experiments}). 

In fact, it turns out that a more correct description of the same diffraction pattern can be achieved within the original space group, provided one introduces two different wave vectors $\bm{k}_1 = (0, \frac{3}{4}, 0)$ and $\bm{k}_2 = (0, \frac{1}{4}, 0)$. It is noteworthy that $\bm{k}_2 = 3\bm{k}_1 - \bm{\tau}_{(0,2,0)}$, where $\bm{\tau}_{(0,2,0)}$ is a reciprocal lattice vector, and thus can be regarded as the third harmonic of $\bm{k}_1$. Its existence implies that significant squaring of the structure takes place.
\footnote{Strictly speaking, the term ``squaring'' applies to a non-sinusoidal periodic wave form and, by extension, to a magnetic structure consisting of $n$-up--$n$-down moment sequences. Here we use it in the sense of equal magnetic moments reaching full saturation value at $T \ll \TN$.} 
In the following, the notations $hkl^{+{\slash}-}$ and $hkl^{+{\slash}-3}$ represent satellites associated with $\bm{k}_1$ and $\bm{k}_2$ (3rd harmonic), respectively (see, e.g.,  Fig.~\ref{diff_patt_1p7}). 

\begin{figure}  
	\centering
	\includegraphics [width=0.97\columnwidth, angle=0] {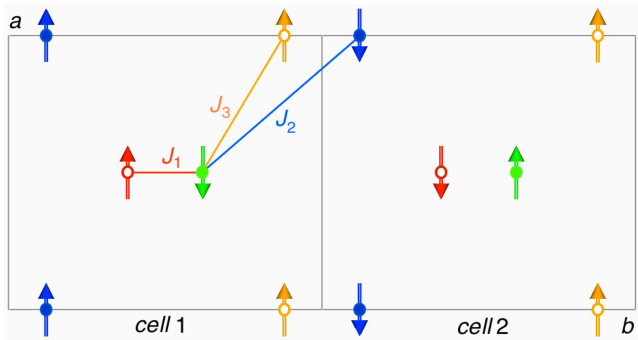}
	\caption{\label{structure}(Color online) Schematic representation of the fully squared magnetic structure yielding the best Rietveld refinement of the NPD data at $T = 1.7$ K (see text). Arrows parallel to the $a$ axis represent Nd magnetic moments located in the z = 1/4 (blue, green : closed circles) and z = 3/4 (red, orange : open circles) planes, respectively. Moments in cells 3 and 4 (not shown) are antiparallel to those in cells 1 and 2, respectively. $J_1$, $J_2$, and $J_3$ denote exchange interactions between Nd near neighbors, as discussed in Sect.~\ref{s:discus}.}
\end{figure}

As a first step, since the diffraction pattern was measured significantly below \TN, we have tentatively assumed that the structure consists of fully saturated Nd magnetic moments of \textit{equal size}. In the primitive (single-$\bm{k}$) representation, it is easy to see that the wave vector $\bm{k}_P = (0, \frac{1}{4}, 0)$ implies four ``\hbox{+\ +\ --\ --}'' sequences propagating along $b$ on four independent $P$ sublattices. We thus have carried out a systematic investigation of all possible phase relationships between the four magnetic sublattices that produce constant moments at the Nd sites and no net magnetization. From a total of 256, only 24 correspond to distinct magnetic structures, for which \textsc{FullProf} refinements have been performed. The best fit, represented by the solid line in Fig.~\ref{diff_patt_1p7}, has a magnetic $R$-factor of 19 (24) excluding (including) the contaminated regions denoted by stars in Fig.~\ref{diff_patt_1p7}, and corresponds to the collinear structure represented in Fig.~\ref{structure}.  The refinement is very sensitive to  the orientation of the moments, which are undoubtedly parallel to the easy $a$ axis, as was inferred earlier from the bulk magnetization measurements.\cite{Kunimori'12} The magnetic $R$-factor for the next best solution is 33 (36), indicating that the agreement with the experimental data is significantly worse. In particular, it completely fails to account for the strong intensity of the 010$^-$ peak near $2\theta = 10\dg$. 
 
Returning to the two-$\bm{k}$ description, it can first be noted that the Fourier representation of constant-moment structures with a periodicity of 4 unit cells along the $b$ direction necessarily involves the two \V{k} vectors $\bm{k}_1$ and $\bm{k}_2$. The solution obtained above can indeed be reproduced using Fourier components $f_1 = 2.39 \mub$ and $f_2 = 0.99 \mub$ with appropriate phases, namely
 
 \begin{equation}\label{eqn1_whole}
\begin {split}
\bm{m}_{l,1} & = \bm{R}_{\bm{k}_1}\cos2\pi[\bm{k_1}\cdot\bm{R}_l+15/16]\\
& \quad +\bm{R}_{\bm{k}_2}\cos2\pi[\bm{k_2}\cdot\bm{R}_l+5/16]\\
\bm{m}_{l,2} & = \bm{R}_{\bm{k}_1}\cos2\pi[\bm{k_1}\cdot\bm{R}_l+13/16]\\
&\quad +\bm{R}_{\bm{k}_2}\cos2\pi[\bm{k_2}\cdot\bm{R}_l+15/16]
\end{split}
\end{equation}

\noindent
for the Nd atoms lying in the $z = 1/4$ and $z = 3/4$ planes, respectively. Here $\bm{R}_l$ represents the coordinates of the $l$th unit cell, and $\bm{R}_{\bm{k}_i} = (0, f_i, 0)$, $i = \{1,2\}$ are the (real) vector Fourier components associated with $\bm{k}_1$ and $\bm{k}_2$. In this solution, full squaring is ensured by keeping the phases as above, and the ratio $f_2/f_1$ equal to $\sqrt{2}-1$. Under these conditions, the refinement obtained is the same as in the equivalent primitive description, displayed in Fig.~\ref{diff_patt_1p7}, with Nd moments equal to 1.82(2) \mub. The only notable discrepancy (apart from the contaminations near $2\theta = 35$ and 40\dg, which already exist above \TN\ and up to RT) is some overestimation of the intensity of the 001$^{\pm 3}$ magnetic satellite near 15.5\dg (light-green trace in the figure).  The $b^{\ast}$ component of the $\bm{k}_1$ vector refines to 0.750(1) r.l.u., i.e. to 3/4 within experimental accuracy.

If the ``full-squaring'' condition is relaxed, the agreement with the measured data slightly improves for $f_1 = 2.42 \mub$ and $f_2 = 0.81 \mub$ ($f_2/f_1 = 0.34 < \sqrt{2}-1$), mainly regarding the intensity of the 001$^{\pm 3}$ peak. This points to an incomplete squaring of the structure. Assuming (somewhat arbitrarily\footnote{the global phase cannot be deduced from a NPD experiment, which measures only the squared module of the magnetic structure factor.}) 
that the  phases remain the same as for the constant-moment solution, one gets two slightly different values, 1.68 and 1.92\mub, for the moments at the Nd sites.

\begin{figure}  
	\centering
	\includegraphics [width=0.97\columnwidth, angle=0] {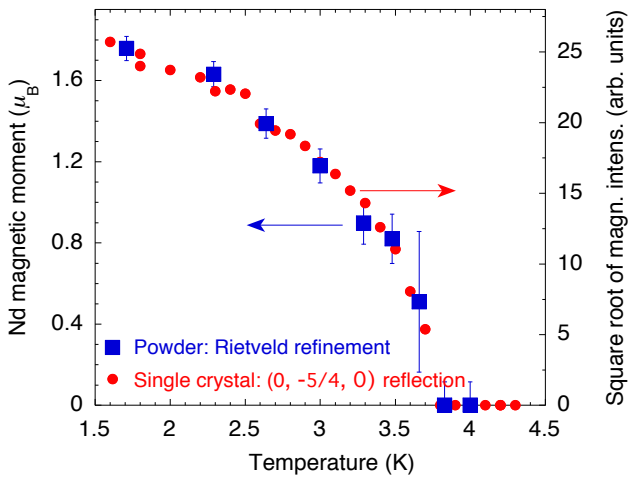}
	\caption{\label{int(t)}(Color online)  Temperature dependence of the magnetic diffraction component in \NdFeAl. Closed squares (left scale): Nd magnetic moment obtained from the Rietveld refinements of the NPD patterns using the magnetic structure described in the text. Closed circles (right scale): square root of the integrated intensity of the $0\bar{2}0^{+}$ satellite [$\bm{Q}=(0, -\frac{5}{4},0)$]. The vertical scales have been adjusted to emphasize the similarity of the temperature variations.}
\end{figure}

With increasing temperature, the magnetic signal decreases gradually, but no change is observed in the position of the peaks. At $T = 2.7$ K, the refinement yields the same value $k_1 = 0.749(1)$, as at 1.7 K. The temperature dependence plotted in Fig.~\ref{int(t)} shows that the Nd magnetic moment vanishes at the Néel temperature $\TN = 3.9$ K. No anomaly is observed near the temperature of 2--2.5 K, at which anomalies were reported from previous experiments.\cite{Kunimori'12, Adroja'13rev,Tanida'priv} Neither could any conclusive change in the relative intensities of the two Fourier components be ascertained, within experimental accuracy, because of the overall loss of magnetic intensity in the temperature range of interest.

\subsection{Magnetic structure: single-crystal diffraction}

\begin{figure}  
	\centering
	\includegraphics [width=0.70\columnwidth, angle=0] {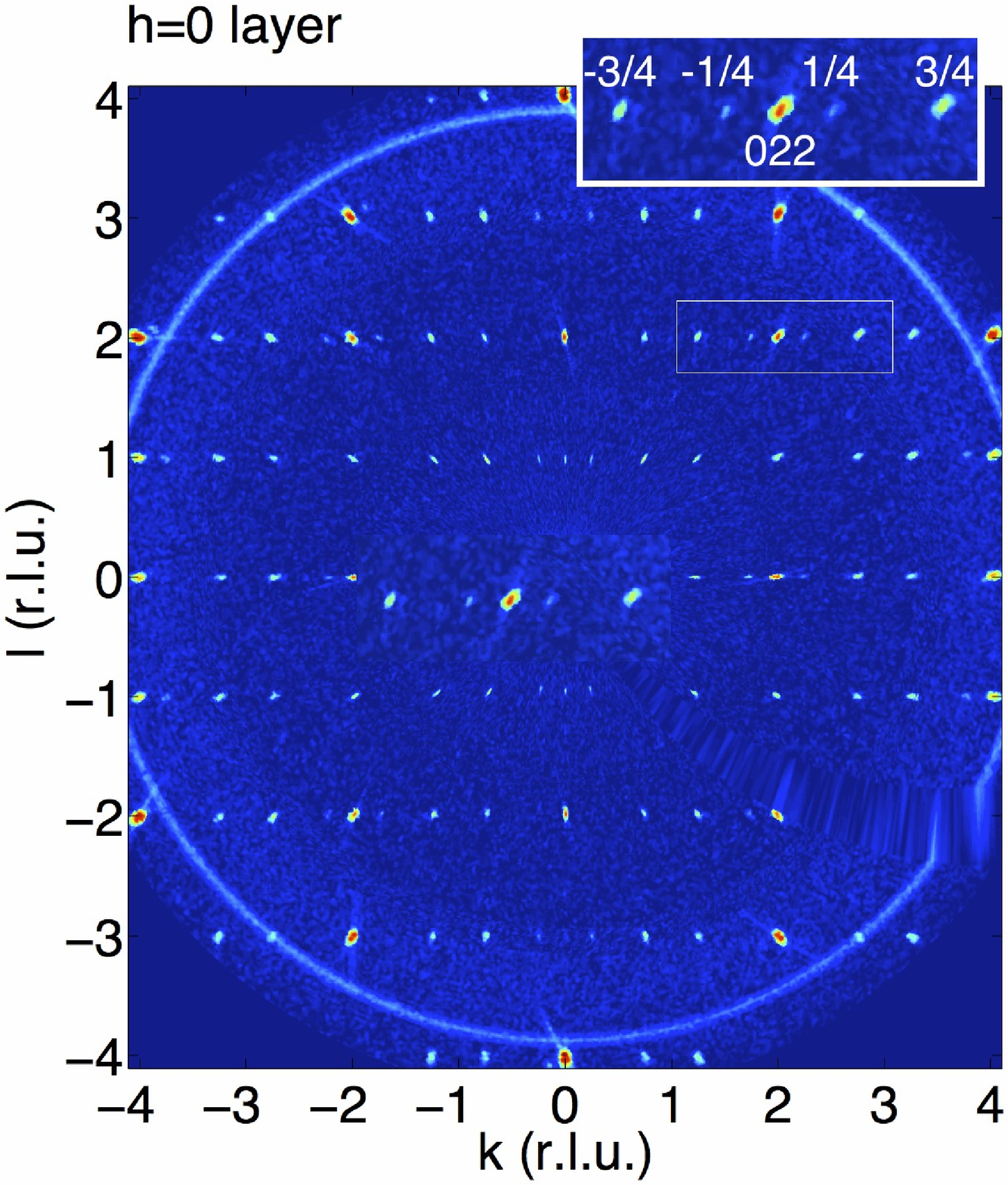}
	\vspace {4pt}
		\includegraphics [width=0.70\columnwidth, angle=0] {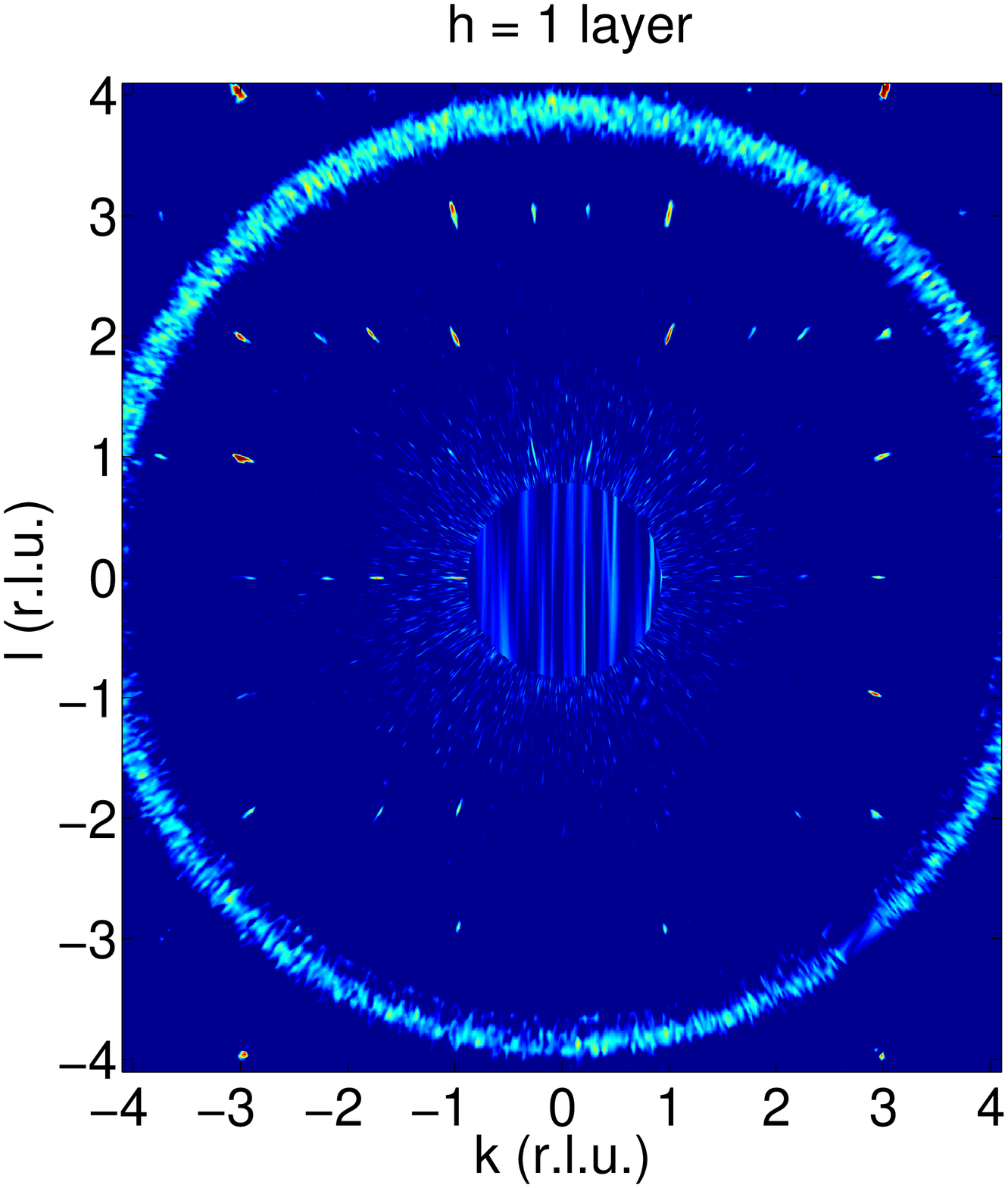}
	\vspace {4pt}
		\includegraphics [width=0.70\columnwidth, angle=0] {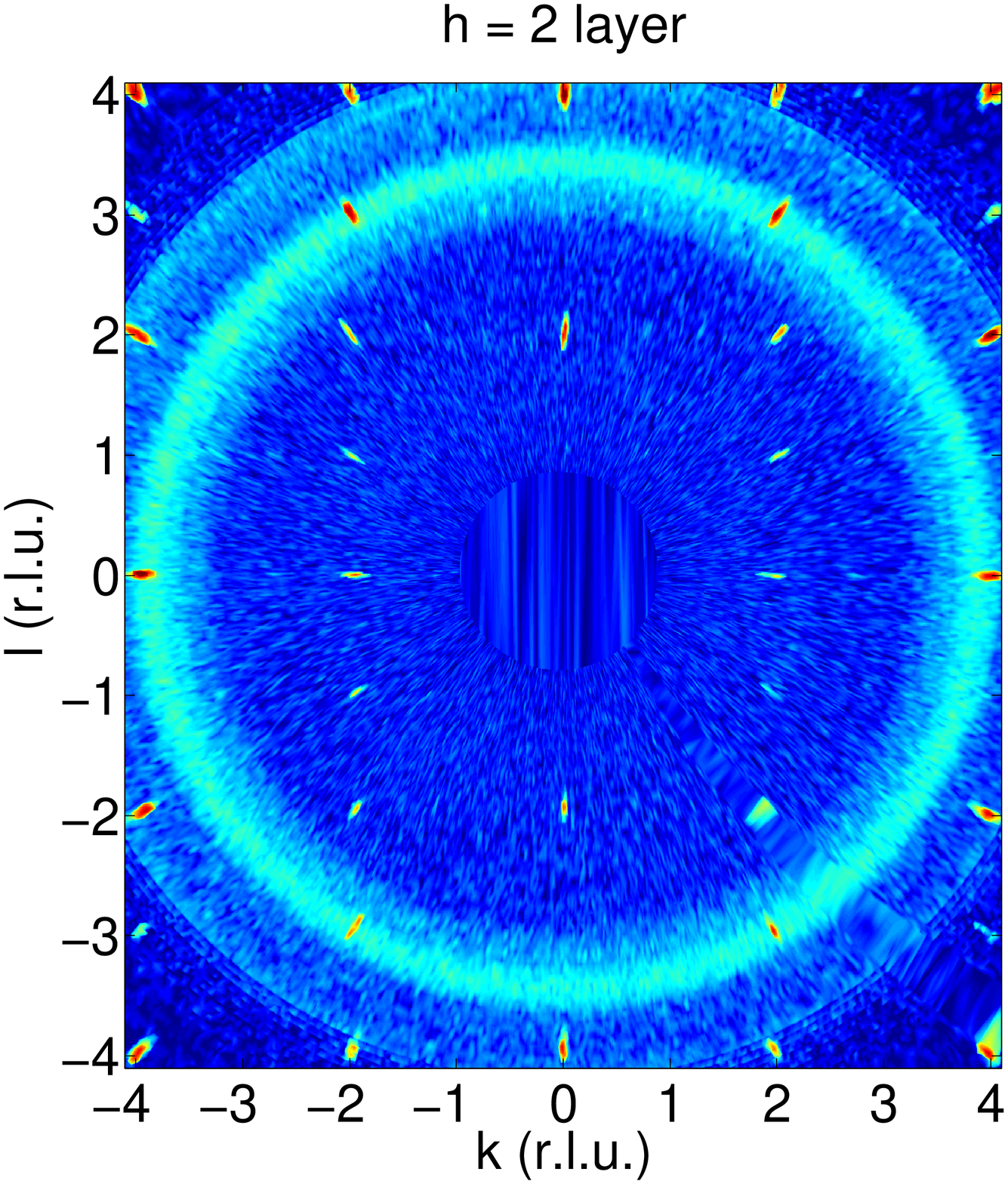}
	\caption{\label{maps}(Color online) Diffraction maps on \NdFeAl\ in the $(0,k,l)$ (upper frame), $(1,k,l)$ (middle frame), and $(2,k,l)$ (lower frame) planes in reciprocal space, obtained from the PSD measurements on Super-6T2. The incident neutron wavelength was $\lambda_i = 2.345$~\AA\ (PG monochromator) for $h=0$ and $\lambda_i = 0.902$~\AA\ (Cu monochromator) for $h=1$ and 2.}
\end{figure}

Single-crystal experiments have been performed on 6T2 using the PSD at an incident neutron wavelength $\lambda_i = 2.345$~\AA. Figure~\ref{maps}(a) shows a cut in the $h=0$ plane through the data measured at $T =  1.7$ K. The nuclear peaks are observed at the positions ($k$ even, $l$ even if $k=0$) corresponding to the reflection conditions for the $Cmcm$ space group, namely: $h+k=2n$, with $h+l=2n$ if $k$ = 0. The violation observed for $00\bar{1}$ is ascribed to multiple scattering. Satellites are clearly visible at positions shifted by $\pm \frac{1}{4} \bm{b}^{\ast}$ from nuclear peaks. However, even stronger magnetic reflections exist near extinct positions (0, $2n+1 \pm 1/4, l)$. As mentioned in Sect.~\ref{ss:magstrucpowd}, these satellites can be indexed either by artificially lowering the symmetry to primitive orthorhombic (with four independent $oP$ Nd Bravais lattices) or, more correctly, staying in the $Cmcm$ space group (two $oC$ Nd Bravais lattices), by introducing the extra wave vector $\bm{k}_1 = (0, \frac{3}{4}, 0)$. 

To study the higher layers $h=1$ and 2, it was necessary to use a shorter wavelength $\lambda_i = 0.902$~\AA\ from the Cu monochromator so as  to remain within the vertical opening of the cryomagnet and the solid angle covered by the detector. The maps measured at $T=1.7$ K for $h=1$ and 2 are shown in Figs.~\ref{maps}(b) and (c) (for $h=0$, the results are consistent with those of Fig.~\ref{maps}(a), but only the stronger ($\bm{k}_1$) satellites can be detected---see Fig. S1 of the Supplemental Material\cite{Supplemental}). The map for $h=2$ is qualitatively similar to that for  $h=0$, but few satellites remain observable. For $h=1$, nuclear reflection conditions reduce to ``$k$ odd''. Again, the stronger satellites occur near forbidden ($k=2n$) reciprocal lattice points, as implied by the $\bm{k}_1$ wave vector. These observations are fully consistent with the magnetic structure suggested above from the NPD results.

 \begin{figure}  
	\centering
	\includegraphics [width=0.75\columnwidth, angle=0] {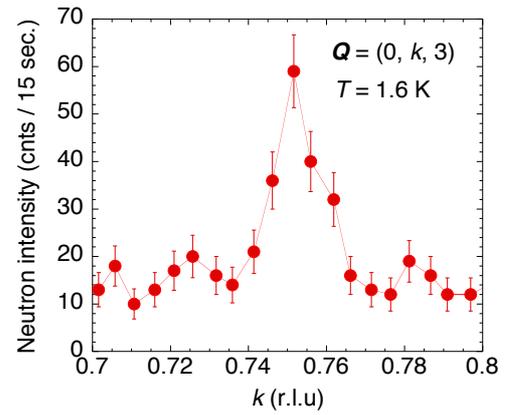}
	\caption{\label{kscan}(Color online) Longitudinal profile of the $000^{+}$ satellite [$\bm{Q}=(0,\frac{3}{4},0)$] measured at $T=1.6$ K.}
\end{figure}
	
 \begin{figure}  
	\centering
	\includegraphics [width=0.75\columnwidth, angle=0] {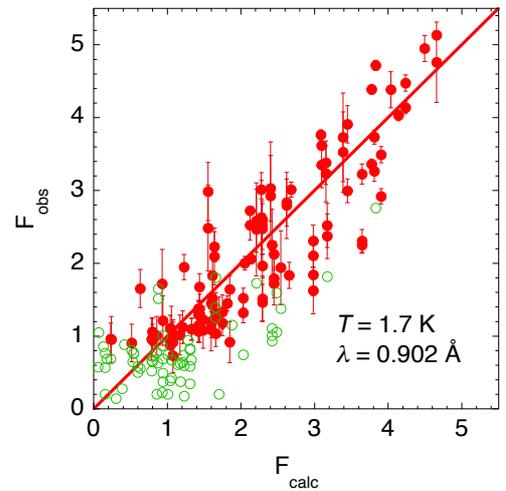}
	\caption{\label{singcrysraf}(Color online) Linear correlation between the experimental magnetic structure factors (square root of the measured integrated intensities divided by the Lorentz factor) for 178 magnetic reflections and the values calculated for the magnetic structure described in the text and shown in Fig.~\ref{structure} (component perpendicular to \V{Q}, corrected for extinction). Closed red (open green) markers denote reflections whose measured $F_{\mathrm{obs}}$ are larger (smaller) than 3 times their standard deviations.}
\end{figure}

Measurements performed using the lifting counter further confirm the validity of this solution. The (0, $k$, 3) longitudinal scan along the $b^{\ast}$ direction presented in Figure~\ref{kscan} demonstrates that the $b^{\ast}$ component of the \V{k} vector is locked at the commensurate value 1/4. The integrated intensities of 178 nonequivalent magnetic reflections have been collected at the base temperature of 1.7 K by performing rocking curves through the nominal peak positions. Figure~\ref{singcrysraf} shows that there is a good linear correlation between the measured $F_\mathrm{obs}$ (square root of the intensity corrected for the Lorentz factor) and $F_\mathrm{calc}$, the component normal to the scattering vector of the calculated magnetic structure factor (including the magnetic form factor and corrected for extinction effects). The refinement has a reliability factor of 16\% (all reflections included), and yields an ordered magnetic moment of 1.95(7) $\mu_B$ on the Nd ions, slightly larger than that obtained in the NPD measurements. 

In Fig.~\ref{int(t)}, the square root of the integrated intensity of the $(0, -\frac{5}{4}, 0)$ magnetic reflection is shown to follow the temperature dependence of the Nd magnetic moment derived from the Rietveld refinements of the powder data (Sect.~\ref{ss:magstrucpowd}). Despite the much smaller temperature step used here, in comparison with the NPD data (Sect.~\ref{ss:magstrucpowd}), there is still no clear evidence for the second transition expected to occur near 2--2.5 K. 

\begin{figure}  
	\centering
	\includegraphics [width=0.85\columnwidth, angle=0] {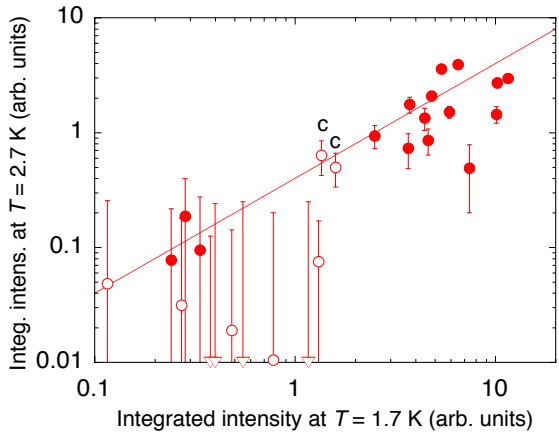}
	\caption{\label{1p6vs2p7k}(Color online)  Comparison of the intensities of selected magnetic reflections at 1.7 and 2.7 K (log-log scales), revealing contrasting behaviors of the $\bm{k}_1$ (closed circles) and $\bm{k}_2$ (open symbols) satellites. Reflections whose intensities drop to less than 0.01 (arb. units) at 2.7 K are denoted by triangles on the horizontal axis. The two reflections $020^{-3}$ and $020^{+3}$ (data points marked by``c'' letters) are thought to contain a contamination since their intensities were subsequently found to retain the same residual intensity in an applied field of 2.6 T, at which other magnetic reflections are fully suppressed (see Section~\ref{ss:magfield}).}
\end{figure}

We have performed a data collection at $T = 2.7$ K on a limited number of representative magnetic reflections associated with the two wave vectors $\bm{k}_1$ and $\bm{k}_2$ (see Fig. S2 of the Supplemental Material\cite{Supplemental}). Their integrated intensities are plotted in Fig.~\ref{1p6vs2p7k} as a function of the corresponding values measured at 1.7 K. Different markers denote the data for $\bm{k}_1$ and $\bm{k}_2$. Despite considerable scatter due to the low measured intensities, the results point to a complete suppression of the $\bm{k}_2$ satellites, while sizable intensity remains in the $\bm{k}_1$ satellites. This observation supports the interpretation of the transition reported previously as being due to a squaring of the magnetic structure. We note that a $\bm{Q}$ scan performed at $T = 2.3$ K through the (0, $k_1$, 3) Bragg peak did not show any deviation of the maximum from the commensurate position $k = 3/4$.
It would now be interesting to trace the temperature dependences of selected reflections from the two subsets in the temperature range of interest.

\subsection{\label{ss:magfield}Magnetic field effects}

\begin{figure}  
	\centering
	\includegraphics [width=0.80\columnwidth, angle=0] {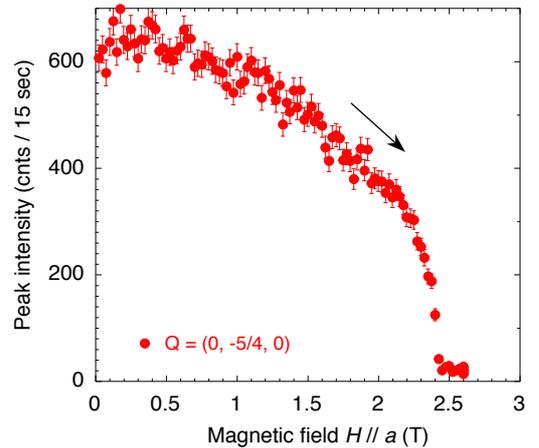}
	\caption{\label{h=0vs2p6t}(Color online) Variation of the peak intensity of the $0\bar{2}0^+$ magnetic satellite [$\bm{Q}=(0, -\frac{5}{4},0)$] in a magnetic field $H \parallel a$ at $T = 1.6$ K.}
\end{figure}

Single-counter measurements have been performed in magnetic fields of up to $H =  2.6$ T applied along the $a$ axis (parallel to the ordered Nd moments). The peak intensity of the $0\bar{2}0^+$ magnetic satellite ($\bm{Q}=(0, -\frac{5}{4},0)$) was traced in an increasing magnetic field at $T = 1.7$ K (Fig.~\ref{h=0vs2p6t}). One observes a steady decrease in the intensity, which drops to zero at about 2.45 T. In their magnetization measurements, Kunimori \etal,\cite{Kunimori'12} observed a steplike increase in $M(H)$ at $T = 1.4$ K for $H \parallel a$, at a transition field of $H_c = 2.45$ T, in excellent agreement with the present data. 

Integrated intensities were measured for a limited set of magnetic reflections in a field of 2.6 T. No residual intensity was detected, except for the 
($020^{-3}$)- and ($020^{+3}$)-type reflections, which exhibit the same contamination as in the corresponding zero-field scans at $T = 2$ K (Fig.~\ref{1p6vs2p7k} and Figs. S3 and S4 of the Supplemental Material\cite{Supplemental}). No evidence was thus found for an intermediate ferrimagnetic phase of the type reported for \TbFeAl.\cite{Reehuis'03}.

\section{\label{s:discus}Discussion and conclusion}

 \begin{figure*}  
	\centering
	\includegraphics [width=0.75\linewidth,angle=0] {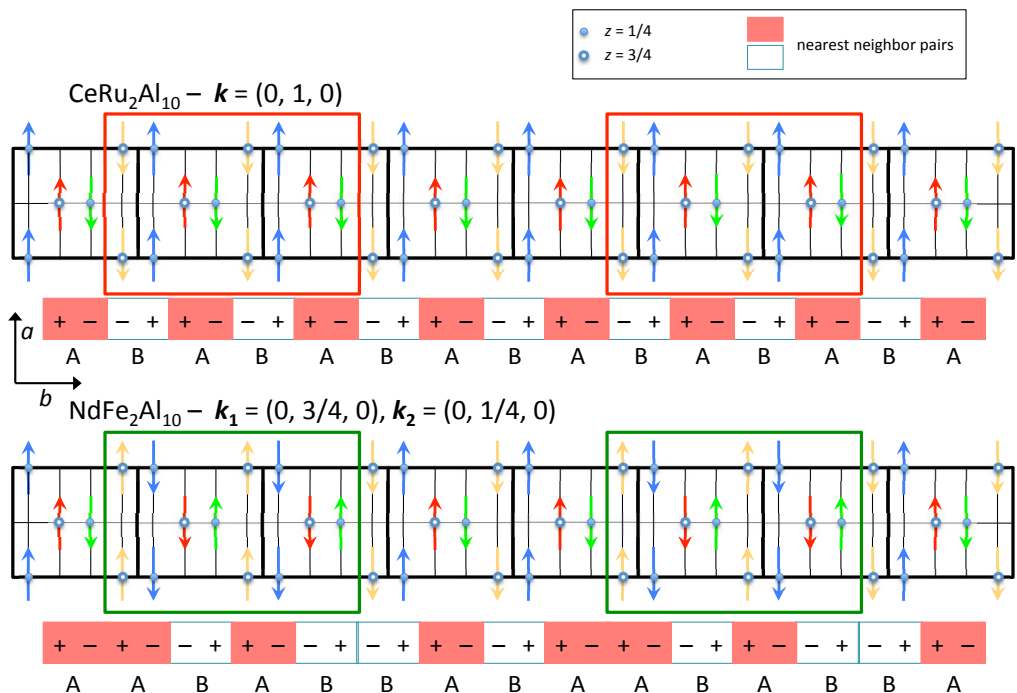}
	\caption{\label{cevsnd}(Color online)  Schematic representation of the magnetic structures in \CeRuAl (above) and \NdFeAl (below). The framed areas denote regions in the \CeRuAl\ lattice whose magnetic moment directions are reversed in \NdFeAl\ (see text). To ease comparison, the Ce magnetic moments are represented tilted from their actual $c$-axis orientation into the plane of the drawing ($a$ axis).}
\end{figure*}

In the previous sections, we have shown that the ordered magnetic structures formed in \NdFeAl\ below $T_N = 3.9$ K can be described, in the crystallographic space group $Cmcm$, using a double-$\bm{k}$ representation with $\bm{k}_1 = (0, \frac{3}{4}, 0)$ and $\bm{k}_2 = (0, \frac{1}{4}, 0)$. At the base temperature $T_{\mathrm{min}} \approx $ 1.6 - 1.7 K, the powder and single-crystal data are consistently accounted for by assuming a collinear ($\bm{m}_{\textrm{Nd}} \parallel a$) structure with constant magnetic moments of $\approx 1.82(2) \mub$ (NPD) or 1.95(7) $\mu_B$ (single-crystal diffraction), in good agreement with the value of $1.91 \mub$ calculated for the Nd$^{3+}$ CF doublet ground state derived in Ref.~\citen{Kunimori'12}. This is in contrast with the simple antiferromagnetic order with $\bm{k} = (0, 1, 0)$ reported previously for \CeRuAl\ and \CeOsAl.\cite{Khalyavin'10,Robert'10,Mignot'11,Kato'11}

In intermetallic lanthanide compounds, couplings between $4f$ magnetic moments are usually dominated by RKKY interactions, and the wave vector of the structure is thus sensitive to peculiarities of the Fermi surface, whose consideration is beyond the scope of this work. However, to try and quantify the competition between the 24 different constant-moment structures (Sect.~\ref{ss:magstrucpowd}) consistent with the $\bm{k}_1$ and $\bm{k}_2$ wave vectors, we have calculated the corresponding exchange energies, assuming couplings extending to the third nearest neighbors (Fig.~\ref{structure}). The results show that the structure derived from the refinements is actually that which minimizes the energy ($E = -8 J_1 - 8 J_2 + 8 J_3$) if all three couplings are AFM (taken here as $J > 0$) and satisfy the two conditions $J_1 > 2 J_3$ (1), \textit{and} $J_2 > J_3$ (2). This  seems reasonable since the distance to the first neighbors is much shorter than to the second and third neighbors. 

It is easy to see in Fig.~\ref{structure} that AFM couplings to the first and second neighbors ($J_1$ and $J_2$) can be simultaneously satisfied if the spin sequence (from left to right) is changed to ``+\ +\ --\ --\ +\ +\ --\ --'', i.e. blue and red moments up, yellow and green moments down. This sequence actually corresponds to that observed in \CeRuAl\ and \CeOsAl\ (Fig.~\ref{cevsnd}), except for the Ce moments being oriented parallel to $c$, and can be described in the $Cmcm$ space group by the simple wave vector (0, 1, 0). In this structure, all third-neighbor pairs are ferromagnetic (FM) and, in Ref.~\citen{Robert'12}, magnetic excitation spectra of \CeRuAl\ in the AFM state could be reproduced assuming exchange interactions with $J_1$ (predominant and anisotropic) and $J_2$ AFM, and $J_3$ FM. It can be noted that, in the present (\YbFeAl-type\cite{Niemann'95}) structure, there is one transition-metal atom located between third-neighbor rare-earth ions, which may explain why $J_3$ can change significantly from one compound to another.

The structure determined above for \NdFeAl\ suggests that $J_3$ is now AFM. It can be obtained (Fig.~\ref{cevsnd}) from the $\bm{k} = (0, 1, 0)$ structure by flipping the moments in a block of 8 consecutive magnetic planes. The new sequence is identical to that shown in Fig~\ref{structure}. If one denotes the near-neighbor AFM pairs (forming alternating zigzag chains along $c$) as $A \equiv$ ``+\ --'' and $B \equiv$ ``--\ +'', it is clear that the change from ABABABAB to A$\vert$ABAB$\vert$BAB can be described by the introduction of periodic spin discommensurations (sometimes also termed ``spin-slips''\cite{Cowley'88}). 

At the borders of one flipped block, $J_2$ couplings become frustrated (FM), whereas originally FM third-neighbor pairs become AFM. Correspondingly, the exchange energy $E$ changes from $-8 J_1 - 16 J_2 + 16 J_3$ to $-8 J_1 - 8 J_2 + 8 J_3$, and the spin-block reversal thus results in a more stable structure provided $J_3$ is larger than $J_2$. This, however, is in contradiction to condition (2) above. Therefore, while the comparison of exchange energies provides a hint as to why this particular \{$\bm{k}_1$, $\bm{k}_2$\} structure is favored with respect to others, it cannot explain \textit{simultaneously} why a  double-$\bm{k}$ structure occurs in the first place, rather than $\bm{k} = (0, 1, 0)$ antiferromagnetism existing in \CeRuAl. An improved model, e.g. taking into account more distant neighbors and/or anisotropic couplings, is needed to solve this problem.

The commensurate structure of \NdFeAl\ has strong similarities with that previously observed by Reehuis \etal \cite{Reehuis'00,Reehuis'03} in \TbFeAl\ below 11 K. In the latter compound, the magnetic period along $b$ is equal to 5, rather than 4, unit cells and the order at base temperature was shown to result from the squaring of the sine-wave modulation which forms below \TN. Using the same notation as above, that structure can be written as A$\vert$ABABA$\vert$ABAB, i.e. with a longer interval between discommensurations.\footnote{Another difference due to the different parity is that, unlike in \NdFeAl, the two $oC$ sublattices ($z = 1/4$ or $z = 3/4$) now carry net magnetizations of opposite signs.} Such steplike reversals of the moment direction likely represent the way in which the system achieves the magnetic period prescribed by long-range RKKY interactions while fulfilling local constraints embedded here in the phenomenological constants $J_1$, $J_2$, $J_3$. 

The existence of a second phase transition below \TN, at about 2--2.5 K seems related to the appearance of the magnetic satellites associated with the wave vector $\bm{k}_2 = (0, \frac{1}{4}, 0)$. This wave vector can equivalently be written as $(0, \frac{9}{4}, 0)$ and plays the same role of the third harmonic in squared ``$n$-up--$n$-down'' structures, by allowing the moments at all sites to reach full saturation as $T \to 0$.
Concerning the effect of a magnetic field at the lowest temperature, we find no indication of an intermediate spin-flip transition to a ferrimagnetic phase of the type reported in Ref.~\citen{Reehuis'03} for \TbFeAl. This is consistent with the existence of a single anomaly in the magnetic isotherm reported in Ref.~\citen{Kunimori'12}, in contrast to the two-step $M(H)$ curve in the Tb case.\cite{Reehuis'03} The metamagnetic jump at $H_c = 2.45$ T coincides with the suppression of all measured magnetic satellites in our single-crystal experiments, suggesting that the system directly reaches a fully polarized state.

Finally, it results from the above discussion that the contrast between the complex commensurate order occurring in \NdFeAl\ and the simple AFM one found previously in its cerium-ruthenium and cerium-osmium homologues could be explained by relatively minor changes in the magnitude of the exchange interactions, in particular the coupling  $J_3$ between third neighbors. However, this similarity may be deceiving because the bulk of experimental evidence points to an unconventional mechanism of magnetic interactions in \CeTAl\ compounds, with a prominent role of anisotropic $c$-$f$ hybridization. This should be the subject of future investigations.

\iftoggle{prb}{
\begin{acknowledgments}
}{
\begin{acknowledgment}
}
We are grateful to J.-L. Meuriot, Th. Robillard, J. Dupont, and X. Guillou for technical support during the experiments. K. S. was supported by a \textit{Japan Society for the Promotion of Science} Research Fellowship.
\iftoggle{prb}{
\end{acknowledgments}
}{
\end{acknowledgment}
}

\iftoggle{prb}{
}{
\bibliographystyle{jpsj}
}


\begin{thebibliography}{31}%
\makeatletter
\providecommand \@ifxundefined [1]{%
 \@ifx{#1\undefined}
}%
\providecommand \@ifnum [1]{%
 \ifnum #1\expandafter \@firstoftwo
 \else \expandafter \@secondoftwo
 \fi
}%
\providecommand \@ifx [1]{%
 \ifx #1\expandafter \@firstoftwo
 \else \expandafter \@secondoftwo
 \fi
}%
\providecommand \natexlab [1]{#1}%
\providecommand \enquote  [1]{``#1''}%
\providecommand \bibnamefont  [1]{#1}%
\providecommand \bibfnamefont [1]{#1}%
\providecommand \citenamefont [1]{#1}%
\providecommand \href@noop [0]{\@secondoftwo}%
\providecommand \href [0]{\begingroup \@sanitize@url \@href}%
\providecommand \@href[1]{\@@startlink{#1}\@@href}%
\providecommand \@@href[1]{\endgroup#1\@@endlink}%
\providecommand \@sanitize@url [0]{\catcode `\\12\catcode `\$12\catcode
  `\&12\catcode `\#12\catcode `\^12\catcode `\_12\catcode `\%12\relax}%
\providecommand \@@startlink[1]{}%
\providecommand \@@endlink[0]{}%
\providecommand \url  [0]{\begingroup\@sanitize@url \@url }%
\providecommand \@url [1]{\endgroup\@href {#1}{\urlprefix }}%
\providecommand \urlprefix  [0]{URL }%
\providecommand \Eprint [0]{\href }%
\providecommand \doibase [0]{http://dx.doi.org/}%
\providecommand \selectlanguage [0]{\@gobble}%
\providecommand \bibinfo  [0]{\@secondoftwo}%
\providecommand \bibfield  [0]{\@secondoftwo}%
\providecommand \translation [1]{[#1]}%
\providecommand \BibitemOpen [0]{}%
\providecommand \bibitemStop [0]{}%
\providecommand \bibitemNoStop [0]{.\EOS\space}%
\providecommand \EOS [0]{\spacefactor3000\relax}%
\providecommand \BibitemShut  [1]{\csname bibitem#1\endcsname}%
\let\auto@bib@innerbib\@empty
\bibitem [{\citenamefont {Muro}\ \emph {et~al.}(2009)\citenamefont {Muro},
  \citenamefont {Motoya}, \citenamefont {Saiga},\ and\ \citenamefont
  {Takabatake}}]{Muro'09}%
  \BibitemOpen
  \bibfield  {author} {\bibinfo {author} {\bibfnamefont {Y.}~\bibnamefont
  {Muro}}, \bibinfo {author} {\bibfnamefont {K.}~\bibnamefont {Motoya}},
  \bibinfo {author} {\bibfnamefont {Y.}~\bibnamefont {Saiga}}, \ and\ \bibinfo
  {author} {\bibfnamefont {T.}~\bibnamefont {Takabatake}},\ }\href@noop {}
  {\bibfield  {journal} {\bibinfo  {journal} {J. Phys. Soc. Jpn.}\ }\textbf
  {\bibinfo {volume} {78}},\ \bibinfo {pages} {083707} (\bibinfo {year}
  {2009})}\BibitemShut {NoStop}%
\bibitem [{\citenamefont {Nishioka}\ \emph {et~al.}(2009)\citenamefont
  {Nishioka}, \citenamefont {Kawamura}, \citenamefont {Takesaka}, \citenamefont
  {Kobayashi}, \citenamefont {Kato}, \citenamefont {Matsumura}, \citenamefont
  {Kodama}, \citenamefont {Matsubayashi},\ and\ \citenamefont
  {Uwatoko}}]{Nishioka'09}%
  \BibitemOpen
  \bibfield  {author} {\bibinfo {author} {\bibfnamefont {T.}~\bibnamefont
  {Nishioka}}, \bibinfo {author} {\bibfnamefont {Y.}~\bibnamefont {Kawamura}},
  \bibinfo {author} {\bibfnamefont {T.}~\bibnamefont {Takesaka}}, \bibinfo
  {author} {\bibfnamefont {R.}~\bibnamefont {Kobayashi}}, \bibinfo {author}
  {\bibfnamefont {H.}~\bibnamefont {Kato}}, \bibinfo {author} {\bibfnamefont
  {M.}~\bibnamefont {Matsumura}}, \bibinfo {author} {\bibfnamefont
  {K.}~\bibnamefont {Kodama}}, \bibinfo {author} {\bibfnamefont
  {K.}~\bibnamefont {Matsubayashi}}, \ and\ \bibinfo {author} {\bibfnamefont
  {Y.}~\bibnamefont {Uwatoko}},\ }\href@noop {} {\bibfield  {journal} {\bibinfo
   {journal} {J. Phys. Soc. Jpn.}\ }\textbf {\bibinfo {volume} {78}},\ \bibinfo
  {pages} {123705} (\bibinfo {year} {2009})}\BibitemShut {NoStop}%
\bibitem [{\citenamefont {Strydom}(2009)}]{Strydom'09}%
  \BibitemOpen
  \bibfield  {author} {\bibinfo {author} {\bibfnamefont {A.~M.}\ \bibnamefont
  {Strydom}},\ }\href@noop {} {\bibfield  {journal} {\bibinfo  {journal}
  {Physica\ B}\ }\textbf {\bibinfo {volume} {404}},\ \bibinfo {pages} {2981}
  (\bibinfo {year} {2009})}\BibitemShut {NoStop}%
\bibitem [{\citenamefont {Kondo}\ \emph {et~al.}(2011)\citenamefont {Kondo},
  \citenamefont {Wang}, \citenamefont {Kindo}, \citenamefont {Ogane},
  \citenamefont {Kawamura}, \citenamefont {Tanimoto}, \citenamefont {Nishioka},
  \citenamefont {Tanaka}, \citenamefont {Tanida},\ and\ \citenamefont
  {Sera}}]{Kondo'11}%
  \BibitemOpen
  \bibfield  {author} {\bibinfo {author} {\bibfnamefont {A.}~\bibnamefont
  {Kondo}}, \bibinfo {author} {\bibfnamefont {J.}~\bibnamefont {Wang}},
  \bibinfo {author} {\bibfnamefont {K.}~\bibnamefont {Kindo}}, \bibinfo
  {author} {\bibfnamefont {Y.}~\bibnamefont {Ogane}}, \bibinfo {author}
  {\bibfnamefont {Y.}~\bibnamefont {Kawamura}}, \bibinfo {author}
  {\bibfnamefont {S.}~\bibnamefont {Tanimoto}}, \bibinfo {author}
  {\bibfnamefont {T.}~\bibnamefont {Nishioka}}, \bibinfo {author}
  {\bibfnamefont {D.}~\bibnamefont {Tanaka}}, \bibinfo {author} {\bibfnamefont
  {H.}~\bibnamefont {Tanida}}, \ and\ \bibinfo {author} {\bibfnamefont
  {M.}~\bibnamefont {Sera}},\ }\href@noop {} {\bibfield  {journal} {\bibinfo
  {journal} {Phys. Rev. B}\ }\textbf {\bibinfo {volume} {83}},\ \bibinfo
  {pages} {180415} (\bibinfo {year} {2011})}\BibitemShut {NoStop}%
\bibitem [{\citenamefont {Tanida}\ \emph {et~al.}(2012)\citenamefont {Tanida},
  \citenamefont {Nonaka}, \citenamefont {Tanaka}, \citenamefont {Sera},
  \citenamefont {Kawamura}, \citenamefont {Uwatoko}, \citenamefont {Nishioka},\
  and\ \citenamefont {Matsumura}}]{Tanida'12}%
  \BibitemOpen
  \bibfield  {author} {\bibinfo {author} {\bibfnamefont {H.}~\bibnamefont
  {Tanida}}, \bibinfo {author} {\bibfnamefont {Y.}~\bibnamefont {Nonaka}},
  \bibinfo {author} {\bibfnamefont {D.}~\bibnamefont {Tanaka}}, \bibinfo
  {author} {\bibfnamefont {M.}~\bibnamefont {Sera}}, \bibinfo {author}
  {\bibfnamefont {Y.}~\bibnamefont {Kawamura}}, \bibinfo {author}
  {\bibfnamefont {Y.}~\bibnamefont {Uwatoko}}, \bibinfo {author} {\bibfnamefont
  {T.}~\bibnamefont {Nishioka}}, \ and\ \bibinfo {author} {\bibfnamefont
  {M.}~\bibnamefont {Matsumura}},\ }\href@noop {} {\bibfield  {journal}
  {\bibinfo  {journal} {Phys. Rev. B}\ }\textbf {\bibinfo {volume} {85}},\
  \bibinfo {pages} {205208} (\bibinfo {year} {2012})}\BibitemShut {NoStop}%
\bibitem [{\citenamefont {Kobayashi}\ \emph {et~al.}(2011)\citenamefont
  {Kobayashi}, \citenamefont {Kawamura}, \citenamefont {Nishioka},
  \citenamefont {Kato}, \citenamefont {Matsumura}, \citenamefont {Kodama},
  \citenamefont {Tanida}, \citenamefont {Sera}, \citenamefont {Matsubayashi},\
  and\ \citenamefont {Uwakoto}}]{Kobayashi'11}%
  \BibitemOpen
  \bibfield  {author} {\bibinfo {author} {\bibfnamefont {R.}~\bibnamefont
  {Kobayashi}}, \bibinfo {author} {\bibfnamefont {Y.}~\bibnamefont {Kawamura}},
  \bibinfo {author} {\bibfnamefont {T.}~\bibnamefont {Nishioka}}, \bibinfo
  {author} {\bibfnamefont {H.}~\bibnamefont {Kato}}, \bibinfo {author}
  {\bibfnamefont {M.}~\bibnamefont {Matsumura}}, \bibinfo {author}
  {\bibfnamefont {K.}~\bibnamefont {Kodama}}, \bibinfo {author} {\bibfnamefont
  {H.}~\bibnamefont {Tanida}}, \bibinfo {author} {\bibfnamefont
  {M.}~\bibnamefont {Sera}}, \bibinfo {author} {\bibfnamefont {K.}~\bibnamefont
  {Matsubayashi}}, \ and\ \bibinfo {author} {\bibfnamefont {Y.}~\bibnamefont
  {Uwakoto}},\ }\href@noop {} {\bibfield  {journal} {\bibinfo  {journal} {J.
  Phys. Soc. Jpn. Suppl.}\ }\textbf {\bibinfo {volume} {80SA}},\ \bibinfo
  {pages} {SA044} (\bibinfo {year} {2011})}\BibitemShut {NoStop}%
\bibitem [{\citenamefont {Kunimori}\ \emph {et~al.}(2012)\citenamefont
  {Kunimori}, \citenamefont {Nakamura}, \citenamefont {Nohara}, \citenamefont
  {Tanida}, \citenamefont {Sera}, \citenamefont {Nishioka},\ and\ \citenamefont
  {Matsumura}}]{Kunimori'12}%
  \BibitemOpen
  \bibfield  {author} {\bibinfo {author} {\bibfnamefont {K.}~\bibnamefont
  {Kunimori}}, \bibinfo {author} {\bibfnamefont {M.}~\bibnamefont {Nakamura}},
  \bibinfo {author} {\bibfnamefont {H.}~\bibnamefont {Nohara}}, \bibinfo
  {author} {\bibfnamefont {H.}~\bibnamefont {Tanida}}, \bibinfo {author}
  {\bibfnamefont {M.}~\bibnamefont {Sera}}, \bibinfo {author} {\bibfnamefont
  {T.}~\bibnamefont {Nishioka}}, \ and\ \bibinfo {author} {\bibfnamefont
  {M.}~\bibnamefont {Matsumura}},\ }\href@noop {} {\bibfield  {journal}
  {\bibinfo  {journal} {Phys. Rev. B}\ }\textbf {\bibinfo {volume} {86}},\
  \bibinfo {pages} {245106} (\bibinfo {year} {2012})}\BibitemShut {NoStop}%
\bibitem [{\citenamefont {Robert}\ \emph {et~al.}(2010)\citenamefont {Robert},
  \citenamefont {Mignot}, \citenamefont {Andr{\'e}}, \citenamefont {Nishioka},
  \citenamefont {Kobayashi}, \citenamefont {Matsumura}, \citenamefont {Tanida},
  \citenamefont {Tanaka},\ and\ \citenamefont {Sera}}]{Robert'10}%
  \BibitemOpen
  \bibfield  {author} {\bibinfo {author} {\bibfnamefont {J.}~\bibnamefont
  {Robert}}, \bibinfo {author} {\bibfnamefont {J.-M.}\ \bibnamefont {Mignot}},
  \bibinfo {author} {\bibfnamefont {G.}~\bibnamefont {Andr{\'e}}}, \bibinfo
  {author} {\bibfnamefont {T.}~\bibnamefont {Nishioka}}, \bibinfo {author}
  {\bibfnamefont {R.}~\bibnamefont {Kobayashi}}, \bibinfo {author}
  {\bibfnamefont {M.}~\bibnamefont {Matsumura}}, \bibinfo {author}
  {\bibfnamefont {H.}~\bibnamefont {Tanida}}, \bibinfo {author} {\bibfnamefont
  {D.}~\bibnamefont {Tanaka}}, \ and\ \bibinfo {author} {\bibfnamefont
  {M.}~\bibnamefont {Sera}},\ }\href@noop {} {\bibfield  {journal} {\bibinfo
  {journal} {Phys. Rev. B}\ }\textbf {\bibinfo {volume} {82}},\ \bibinfo
  {pages} {100404(R)} (\bibinfo {year} {2010})}\BibitemShut {NoStop}%
\bibitem [{\citenamefont {Khalyavin}\ \emph {et~al.}(2010)\citenamefont
  {Khalyavin}, \citenamefont {Hillier}, \citenamefont {Adroja}, \citenamefont
  {Strydom}, \citenamefont {Manuel}, \citenamefont {Chapon}, \citenamefont
  {Peratheepan}, \citenamefont {Knight}, \citenamefont {Deen}, \citenamefont
  {Ritter}, \citenamefont {Muro},\ and\ \citenamefont
  {Takabatake}}]{Khalyavin'10}%
  \BibitemOpen
  \bibfield  {author} {\bibinfo {author} {\bibfnamefont {D.~D.}\ \bibnamefont
  {Khalyavin}}, \bibinfo {author} {\bibfnamefont {A.~D.}\ \bibnamefont
  {Hillier}}, \bibinfo {author} {\bibfnamefont {D.~T.}\ \bibnamefont {Adroja}},
  \bibinfo {author} {\bibfnamefont {A.~M.}\ \bibnamefont {Strydom}}, \bibinfo
  {author} {\bibfnamefont {P.}~\bibnamefont {Manuel}}, \bibinfo {author}
  {\bibfnamefont {L.~C.}\ \bibnamefont {Chapon}}, \bibinfo {author}
  {\bibfnamefont {P.}~\bibnamefont {Peratheepan}}, \bibinfo {author}
  {\bibfnamefont {K.}~\bibnamefont {Knight}}, \bibinfo {author} {\bibfnamefont
  {P.}~\bibnamefont {Deen}}, \bibinfo {author} {\bibfnamefont {C.}~\bibnamefont
  {Ritter}}, \bibinfo {author} {\bibfnamefont {Y.}~\bibnamefont {Muro}}, \ and\
  \bibinfo {author} {\bibfnamefont {T.}~\bibnamefont {Takabatake}},\
  }\href@noop {} {\bibfield  {journal} {\bibinfo  {journal} {Phys. Rev. B}\
  }\textbf {\bibinfo {volume} {82}},\ \bibinfo {pages} {100405(R)} (\bibinfo
  {year} {2010})}\BibitemShut {NoStop}%
\bibitem [{\citenamefont {Mignot}\ \emph {et~al.}(2011)\citenamefont {Mignot},
  \citenamefont {Robert}, \citenamefont {Andr{\'e}}, \citenamefont {Bataille},
  \citenamefont {Nishioka}, \citenamefont {Kobayashi}, \citenamefont
  {Matsumura}, \citenamefont {Tanida}, \citenamefont {Tanaka},\ and\
  \citenamefont {Sera}}]{Mignot'11}%
  \BibitemOpen
  \bibfield  {author} {\bibinfo {author} {\bibfnamefont {J.-M.}\ \bibnamefont
  {Mignot}}, \bibinfo {author} {\bibfnamefont {J.}~\bibnamefont {Robert}},
  \bibinfo {author} {\bibfnamefont {G.}~\bibnamefont {Andr{\'e}}}, \bibinfo
  {author} {\bibfnamefont {A.~M.}\ \bibnamefont {Bataille}}, \bibinfo {author}
  {\bibfnamefont {T.}~\bibnamefont {Nishioka}}, \bibinfo {author}
  {\bibfnamefont {R.}~\bibnamefont {Kobayashi}}, \bibinfo {author}
  {\bibfnamefont {M.}~\bibnamefont {Matsumura}}, \bibinfo {author}
  {\bibfnamefont {H.}~\bibnamefont {Tanida}}, \bibinfo {author} {\bibfnamefont
  {D.}~\bibnamefont {Tanaka}}, \ and\ \bibinfo {author} {\bibfnamefont
  {M.}~\bibnamefont {Sera}},\ }\href@noop {} {\bibfield  {journal} {\bibinfo
  {journal} {J. Phys. Soc. Jpn. Suppl.}\ }\textbf {\bibinfo {volume} {80SA}},\
  \bibinfo {pages} {SA022} (\bibinfo {year} {2011})}\BibitemShut {NoStop}%
\bibitem [{\citenamefont {Robert}\ \emph {et~al.}(2012)\citenamefont {Robert},
  \citenamefont {Mignot}, \citenamefont {Petit}, \citenamefont {Steffens},
  \citenamefont {Nishioka}, \citenamefont {Kobayashi}, \citenamefont
  {Matsumura}, \citenamefont {Tanida}, \citenamefont {Tanaka},\ and\
  \citenamefont {Sera}}]{Robert'12}%
  \BibitemOpen
  \bibfield  {author} {\bibinfo {author} {\bibfnamefont {J.}~\bibnamefont
  {Robert}}, \bibinfo {author} {\bibfnamefont {J.-M.}\ \bibnamefont {Mignot}},
  \bibinfo {author} {\bibfnamefont {S.}~\bibnamefont {Petit}}, \bibinfo
  {author} {\bibfnamefont {P.}~\bibnamefont {Steffens}}, \bibinfo {author}
  {\bibfnamefont {T.}~\bibnamefont {Nishioka}}, \bibinfo {author}
  {\bibfnamefont {R.}~\bibnamefont {Kobayashi}}, \bibinfo {author}
  {\bibfnamefont {M.}~\bibnamefont {Matsumura}}, \bibinfo {author}
  {\bibfnamefont {H.}~\bibnamefont {Tanida}}, \bibinfo {author} {\bibfnamefont
  {D.}~\bibnamefont {Tanaka}}, \ and\ \bibinfo {author} {\bibfnamefont
  {M.}~\bibnamefont {Sera}},\ }\href {\doibase 10.1103/PhysRevLett.109.267208}
  {\bibfield  {journal} {\bibinfo  {journal} {Phys. Rev. Lett.}\ }\textbf
  {\bibinfo {volume} {109}},\ \bibinfo {pages} {267208} (\bibinfo {year}
  {2012})}\BibitemShut {NoStop}%
\bibitem [{\citenamefont {Adroja}\ \emph {et~al.}(2013)\citenamefont {Adroja},
  \citenamefont {Hillier}, \citenamefont {Muro}, \citenamefont {Takabatake},
  \citenamefont {Strydom}, \citenamefont {Bhattacharyya}, \citenamefont
  {Daoud-Aladin},\ and\ \citenamefont {Taylor}}]{Adroja'13rev}%
  \BibitemOpen
  \bibfield  {author} {\bibinfo {author} {\bibfnamefont {D.~T.}\ \bibnamefont
  {Adroja}}, \bibinfo {author} {\bibfnamefont {A.~D.}\ \bibnamefont {Hillier}},
  \bibinfo {author} {\bibfnamefont {Y.}~\bibnamefont {Muro}}, \bibinfo {author}
  {\bibfnamefont {T.}~\bibnamefont {Takabatake}}, \bibinfo {author}
  {\bibfnamefont {A.~M.}\ \bibnamefont {Strydom}}, \bibinfo {author}
  {\bibfnamefont {A.}~\bibnamefont {Bhattacharyya}}, \bibinfo {author}
  {\bibfnamefont {A.}~\bibnamefont {Daoud-Aladin}}, \ and\ \bibinfo {author}
  {\bibfnamefont {J.~W.}\ \bibnamefont {Taylor}},\ }\href@noop {} {\bibfield
  {journal} {\bibinfo  {journal} {Physica Scripta}\ }\textbf {\bibinfo {volume}
  {88}},\ \bibinfo {pages} {068505} (\bibinfo {year} {2013})}\BibitemShut
  {NoStop}%
\bibitem [{\citenamefont {Tanida}\ \emph {et~al.}(2014)\citenamefont {Tanida}
  \emph {et~al.}}]{Tanida'priv}%
  \BibitemOpen
  \bibfield  {author} {\bibinfo {author} {\bibfnamefont {H.}~\bibnamefont
  {Tanida}} \emph {et~al.},\ }\href@noop {} {} (\bibinfo {year} {2014}),\
  \bibinfo {note} {unpublished}\BibitemShut {NoStop}%
\bibitem [{\citenamefont {Reehuis}\ \emph {et~al.}(2000)\citenamefont
  {Reehuis}, \citenamefont {Fehrmann}, \citenamefont {Wolff}, \citenamefont
  {Jeitschko},\ and\ \citenamefont {Hofmann}}]{Reehuis'00}%
  \BibitemOpen
  \bibfield  {author} {\bibinfo {author} {\bibfnamefont {M.}~\bibnamefont
  {Reehuis}}, \bibinfo {author} {\bibfnamefont {B.}~\bibnamefont {Fehrmann}},
  \bibinfo {author} {\bibfnamefont {M.~W.}\ \bibnamefont {Wolff}}, \bibinfo
  {author} {\bibfnamefont {W.}~\bibnamefont {Jeitschko}}, \ and\ \bibinfo
  {author} {\bibfnamefont {M.}~\bibnamefont {Hofmann}},\ }\href@noop {}
  {\bibfield  {journal} {\bibinfo  {journal} {Physica B: Condensed Matter}\
  }\textbf {\bibinfo {volume} {276-278}},\ \bibinfo {pages} {594} (\bibinfo
  {year} {2000})}\BibitemShut {NoStop}%
\bibitem [{\citenamefont {Reehuis}\ \emph {et~al.}(2003)\citenamefont
  {Reehuis}, \citenamefont {Wolff}, \citenamefont {Krimmel}, \citenamefont
  {Scheidt}, \citenamefont {St\"usser}, \citenamefont {Loidl},\ and\
  \citenamefont {Jeitschko}}]{Reehuis'03}%
  \BibitemOpen
  \bibfield  {author} {\bibinfo {author} {\bibfnamefont {M.}~\bibnamefont
  {Reehuis}}, \bibinfo {author} {\bibfnamefont {M.~W.}\ \bibnamefont {Wolff}},
  \bibinfo {author} {\bibfnamefont {A.}~\bibnamefont {Krimmel}}, \bibinfo
  {author} {\bibfnamefont {E.-W.}\ \bibnamefont {Scheidt}}, \bibinfo {author}
  {\bibfnamefont {N.}~\bibnamefont {St\"usser}}, \bibinfo {author}
  {\bibfnamefont {A.}~\bibnamefont {Loidl}}, \ and\ \bibinfo {author}
  {\bibfnamefont {W.}~\bibnamefont {Jeitschko}},\ }\href@noop {} {\bibfield
  {journal} {\bibinfo  {journal} {Journal of Physics: Condensed Matter}\
  }\textbf {\bibinfo {volume} {15}},\ \bibinfo {pages} {1773} (\bibinfo {year}
  {2003})}\BibitemShut {NoStop}%
\bibitem [{\citenamefont {Thiede}\ \emph {et~al.}(1998)\citenamefont {Thiede},
  \citenamefont {Ebel},\ and\ \citenamefont {Jeitschko}}]{Thiede'98}%
  \BibitemOpen
  \bibfield  {author} {\bibinfo {author} {\bibfnamefont {V.~M.~T.}\
  \bibnamefont {Thiede}}, \bibinfo {author} {\bibfnamefont {T.}~\bibnamefont
  {Ebel}}, \ and\ \bibinfo {author} {\bibfnamefont {W.}~\bibnamefont
  {Jeitschko}},\ }\href@noop {} {\bibfield  {journal} {\bibinfo  {journal} {J.
  Mater. Chem.}\ }\textbf {\bibinfo {volume} {8}},\ \bibinfo {pages} {125}
  (\bibinfo {year} {1998})}\BibitemShut {NoStop}%
\bibitem [{\citenamefont {Sera}\ \emph
  {et~al.}(2013{\natexlab{a}})\citenamefont {Sera}, \citenamefont {Tanaka},
  \citenamefont {Tanida}, \citenamefont {Moriyoshi}, \citenamefont {Ogawa},
  \citenamefont {Kuroiwa}, \citenamefont {Nishioka}, \citenamefont {Matsumura},
  \citenamefont {Kim}, \citenamefont {Tsuji},\ and\ \citenamefont
  {Takata}}]{Sera'13}%
  \BibitemOpen
  \bibfield  {author} {\bibinfo {author} {\bibfnamefont {M.}~\bibnamefont
  {Sera}}, \bibinfo {author} {\bibfnamefont {D.}~\bibnamefont {Tanaka}},
  \bibinfo {author} {\bibfnamefont {H.}~\bibnamefont {Tanida}}, \bibinfo
  {author} {\bibfnamefont {C.}~\bibnamefont {Moriyoshi}}, \bibinfo {author}
  {\bibfnamefont {M.}~\bibnamefont {Ogawa}}, \bibinfo {author} {\bibfnamefont
  {Y.}~\bibnamefont {Kuroiwa}}, \bibinfo {author} {\bibfnamefont
  {T.}~\bibnamefont {Nishioka}}, \bibinfo {author} {\bibfnamefont
  {M.}~\bibnamefont {Matsumura}}, \bibinfo {author} {\bibfnamefont
  {J.}~\bibnamefont {Kim}}, \bibinfo {author} {\bibfnamefont {S.}~\bibnamefont
  {Tsuji}}, \ and\ \bibinfo {author} {\bibfnamefont {M.}~\bibnamefont
  {Takata}},\ }\href@noop {} {\bibfield  {journal} {\bibinfo  {journal} {J.
  Phys. Soc. Jpn.}\ }\textbf {\bibinfo {volume} {82}},\ \bibinfo {pages}
  {024603} (\bibinfo {year} {2013}{\natexlab{a}})}\BibitemShut {NoStop}%
\bibitem [{\citenamefont {Sera}\ \emph
  {et~al.}(2013{\natexlab{b}})\citenamefont {Sera} \emph {et~al.}}]{Sera'priv}%
  \BibitemOpen
  \bibfield  {author} {\bibinfo {author} {\bibfnamefont {M.}~\bibnamefont
  {Sera}} \emph {et~al.},\ }\href@noop {} {} (\bibinfo {year}
  {2013}{\natexlab{b}}),\ \bibinfo {note} {unpublished}\BibitemShut {NoStop}%
\bibitem [{\citenamefont {Rodriguez-Carvajal}(1993)}]{fullprof'93}%
  \BibitemOpen
  \bibfield  {author} {\bibinfo {author} {\bibfnamefont {J.}~\bibnamefont
  {Rodriguez-Carvajal}},\ }\href@noop {} {\bibfield  {journal} {\bibinfo
  {journal} {Physica\ B}\ }\textbf {\bibinfo {volume} {192}},\ \bibinfo {pages}
  {55} (\bibinfo {year} {1993})}\BibitemShut {NoStop}%
\bibitem [{\citenamefont {Rodriguez-Carvajal}(2001)}]{fullprof'01}%
  \BibitemOpen
  \bibfield  {author} {\bibinfo {author} {\bibfnamefont {J.}~\bibnamefont
  {Rodriguez-Carvajal}},\ }\href@noop {} {\bibfield  {journal} {\bibinfo
  {journal} {Commission on Powder Diffraction (IUCr) Newsletter}\ }\textbf
  {\bibinfo {volume} {26}},\ \bibinfo {pages} {12} (\bibinfo {year}
  {2001})}\BibitemShut {NoStop}%
\bibitem [{\citenamefont {Sears}(1992)}]{Sears'92}%
  \BibitemOpen
  \bibfield  {author} {\bibinfo {author} {\bibfnamefont {V.~F.}\ \bibnamefont
  {Sears}},\ }\href@noop {} {\bibfield  {journal} {\bibinfo  {journal} {Neutron
  News}\ }\textbf {\bibinfo {volume} {3}},\ \bibinfo {pages} {26} (\bibinfo
  {year} {1992})}\BibitemShut {NoStop}%
\bibitem [{\citenamefont {Freeman}\ and\ \citenamefont
  {Desclaux}(1979)}]{Freeman'79}%
  \BibitemOpen
  \bibfield  {author} {\bibinfo {author} {\bibfnamefont {A.}~\bibnamefont
  {Freeman}}\ and\ \bibinfo {author} {\bibfnamefont {J.}~\bibnamefont
  {Desclaux}},\ }\href@noop {} {\bibfield  {journal} {\bibinfo  {journal} {J.\
  Magn.\ Magn.\ Mater.}\ }\textbf {\bibinfo {volume} {12}},\ \bibinfo {pages}
  {11} (\bibinfo {year} {1979})}\BibitemShut {NoStop}%
\bibitem [{\citenamefont {Gukasov}\ \emph {et~al.}(2007)\citenamefont
  {Gukasov}, \citenamefont {Goujon}, \citenamefont {Meuriot}, \citenamefont
  {Person}, \citenamefont {Exil},\ and\ \citenamefont {Koskas}}]{Gukasov'07}%
  \BibitemOpen
  \bibfield  {author} {\bibinfo {author} {\bibfnamefont {A.}~\bibnamefont
  {Gukasov}}, \bibinfo {author} {\bibfnamefont {A.}~\bibnamefont {Goujon}},
  \bibinfo {author} {\bibfnamefont {J.-L.}\ \bibnamefont {Meuriot}}, \bibinfo
  {author} {\bibfnamefont {C.}~\bibnamefont {Person}}, \bibinfo {author}
  {\bibfnamefont {G.}~\bibnamefont {Exil}}, \ and\ \bibinfo {author}
  {\bibfnamefont {G.}~\bibnamefont {Koskas}},\ }\href@noop {} {\bibfield
  {journal} {\bibinfo  {journal} {Physica B: Condensed Matter}\ }\textbf
  {\bibinfo {volume} {397}},\ \bibinfo {pages} {131} (\bibinfo {year}
  {2007})}\BibitemShut {NoStop}%
\bibitem [{ccs()}]{ccsl'93}%
  \BibitemOpen
  \href@noop {} {}\bibinfo {note} {J. Brown and J. Matthewman, Cambridge
  Crystallography Subroutine Library, Report No. RAL93-009 (1993)}\BibitemShut
  {NoStop}%
\bibitem [{Note1()}]{Note1}%
  \BibitemOpen
  \bibinfo {note} {Strictly speaking, the term ``squaring'' applies to a
  non-sinusoidal periodic wave form and, by extension, to a magnetic structure
  consisting of $n$-up--$n$-down moment sequences. Here we use it in the sense
  of equal magnetic moments reaching full saturation value at $T \ll \protect
  \ensuremath {T_{N}}$.}\BibitemShut {Stop}%
\bibitem [{Note2()}]{Note2}%
  \BibitemOpen
  \bibinfo {note} {The global phase cannot be deduced from a NPD experiment,
  which measures only the squared module of the magnetic structure
  factor.}\BibitemShut {Stop}%
\bibitem [{Sup()}]{Supplemental}%
  \BibitemOpen
  \href@noop {} {}\bibinfo {note} {See Supplemental Material at [URL] for a
  presentation of additional data sets (single-crystal intensity map, rocking
  curves through magnetic peaks at different temperatures and magnetic
  fields).}\BibitemShut {Stop}%
\bibitem [{\citenamefont {Kato}\ \emph {et~al.}(2011)\citenamefont {Kato},
  \citenamefont {Kobayashi}, \citenamefont {Takesaka}, \citenamefont
  {Nishioka}, \citenamefont {Matsumura}, \citenamefont {Kaneko},\ and\
  \citenamefont {Metoki}}]{Kato'11}%
  \BibitemOpen
  \bibfield  {author} {\bibinfo {author} {\bibfnamefont {H.}~\bibnamefont
  {Kato}}, \bibinfo {author} {\bibfnamefont {R.}~\bibnamefont {Kobayashi}},
  \bibinfo {author} {\bibfnamefont {T.}~\bibnamefont {Takesaka}}, \bibinfo
  {author} {\bibfnamefont {T.}~\bibnamefont {Nishioka}}, \bibinfo {author}
  {\bibfnamefont {M.}~\bibnamefont {Matsumura}}, \bibinfo {author}
  {\bibfnamefont {K.}~\bibnamefont {Kaneko}}, \ and\ \bibinfo {author}
  {\bibfnamefont {N.}~\bibnamefont {Metoki}},\ }\href@noop {} {\bibfield
  {journal} {\bibinfo  {journal} {J. Phys. Soc. Jpn.
  Suppl.}\ }\textbf {\bibinfo {volume} {80}},\ \bibinfo {pages} {073701}
  (\bibinfo {year} {2011})}\BibitemShut {NoStop}%
\bibitem [{\citenamefont {Niemann}\ and\ \citenamefont
  {Jeitschko}(1995)}]{Niemann'95}%
  \BibitemOpen
  \bibfield  {author} {\bibinfo {author} {\bibfnamefont {S.}~\bibnamefont
  {Niemann}}\ and\ \bibinfo {author} {\bibfnamefont {W.}~\bibnamefont
  {Jeitschko}},\ }\href@noop {} {\bibfield  {journal} {\bibinfo  {journal}
  {Z. Kristallogr.}\ }\textbf {\bibinfo {volume} {210}},\
  \bibinfo {pages} {338} (\bibinfo {year} {1995})}\BibitemShut {NoStop}%
\bibitem [{\citenamefont {Cowley}\ and\ \citenamefont
  {Bates}(1988)}]{Cowley'88}%
  \BibitemOpen
  \bibfield  {author} {\bibinfo {author} {\bibfnamefont {R.~A.}\ \bibnamefont
  {Cowley}}\ and\ \bibinfo {author} {\bibfnamefont {S.}~\bibnamefont {Bates}},\
  }\href@noop {} {\bibfield  {journal} {\bibinfo  {journal} {J. Phys. C: Solid
  State Phys.}\ }\textbf {\bibinfo {volume} {21}},\ \bibinfo {pages} {4113}
  (\bibinfo {year} {1988})}\BibitemShut {NoStop}%
\bibitem [{Note3()}]{Note3}%
  \BibitemOpen
  \bibinfo {note} {Another difference due to the different parity is that,
  unlike in NdFe$_{2}$Al$_{10}$, the two $oC$ sublattices ($z = 1/4$ or $z =
  3/4$) now carry net magnetizations of opposite signs.}\BibitemShut {Stop}%
\end{thebibliography}

%

\end{document}